\documentclass[12pt,preprint]{aastex}
\newcommand{\bld}[1]{\mbox{\boldmath$#1$\unboldmath}}
\newcommand{\pomega}{\varpi}
\def\refnew#1{(\ref{#1})}
\newcommand{\w}{\rm w}

\shorttitle{Planets Eccentricities}

\begin{document}

\title{Eccentricity Evolution for Planets in Gaseous Disks}

\author{Peter Goldreich and Re'em Sari}
\affil{130-33 Caltech, Pasadena, CA 91125}

\begin{abstract} 

At least several percent of solar type stars possess giant planets. 
Surprisingly, most move on orbits of substantial eccentricity. We
investigate the hypothesis that interactions between a giant planet 
and the disk from which it forms promote eccentricity growth.
These interactions are concentrated at discrete Lindblad and corotation 
resonances. Interactions at principal Lindblad resonances cause the planet's 
orbit to migrate and open a gap in the disk if the planet is sufficiently 
massive. Those at first order Lindblad and corotation resonances change the 
planet's orbital eccentricity. Eccentricity is excited by interactions at 
external Lindblad resonances which are located on the opposite side of
corotation from the planet, and damped by co-orbital Lindblad
resonances which overlap the planet's orbit. If the planet clears a gap in the 
disk, the rate of eccentricity damping by co-orbital Lindblad resonances is 
reduced. Density gradients associated with the gap activate eccentricity 
damping by corotation resonances at a rate which initially marginally exceeds 
that of eccentricity excitation by external Lindblad resonances. But the 
corotation torque drives a mass flux which reduces the density gradient near 
the resonance. Sufficient partial saturation of corotation resonances can tip 
the balance in favor of eccentricity excitation. A minimal initial eccentricity 
of a few percent is required to overcome viscous diffusion which acts to 
unsaturate corotation resonances by reestablishing the large scale density 
gradient. Thus eccentricity growth is a finite amplitude instability. Formally,
interactions at the apsidal resonance, which is a special kind of co-orbital 
Lindblad resonance, appears to damp eccentricity faster than external Lindblad 
resonances can excite it. However, apsidal waves have such long wavelengths 
that they do not propagate in protoplanetary disks. This reduces eccentricity 
damping by the apsidal resonance to a modest level. 
\end{abstract}

\section{Introduction}

As of today, there are 77 extrasolar planets listed in the California \& 
Carnegie Planet Search home page http://exoplanets.org/almanac.html.  Almost
all of the 57 planets with periods greater than 20 days move on orbits of
substantial eccentricity, the mean and median being 0.35 and 0.37 respectively.
By contrast, the 13 planets with periods less than 7 days have low eccentricity
orbits, 0.03 mean and 0.02 median, apparently the result of eccentricity
damping associated with tides raised in the planets by their central stars.
The critical period separating small from large eccentricity orbits implies
that extrasolar planets have tidal quality factors, $Q\sim 10^5$, similar to
that of Jupiter.

Solar system planets, with the exception of Mercury, have low eccentricity 
orbits.\footnote{We consider Pluto to be a member of the Kuiper belt and not a   
bone fide planet.} Thus the large orbital eccentricities of extrasolar planets 
came as a surprise. How might they have arisen? Interactions among planets 
have been the focus of most previous suggestions
\citep{RAF96,LIA97,WEM96,FHR01,CFT02}. In systems with isolated planets, these 
involve speculative scenarios in which planets were either ejected from the 
system \citep{RAF96} or merged with the remaining planet \citep{LIA97}. Less
attention has been paid to interactions between planets and the disks from 
which they formed. Analytic theory is not capable of resolving whether these 
interactions can produce eccentricity growth.  However, it can identify crucial 
issues and thus provide guidance for targeted simulations. That is the goal of 
our paper. 

Our investigation applies results derived in studies of satellite interactions 
with planetary rings by \cite{GOT79, GOT80} and later extended by \cite{WAR86} 
and \cite{ART93a, ART93b}. \cite{GOT80, GOT81} conclude that interactions at 
Lindblad resonances excite orbital eccentricity of both narrow rings and their 
shepherd satellites, whereas those at corotation resonances damp it. However, 
the balance between excitation and damping is a precarious one; damping exceeds 
driving by about 4.6\% provided the corotation resonances are unsaturated. 
\cite{WAR86} and \cite{ART93b} point out that interactions at co-orbital 
Lindblad resonances damp its orbital eccentricity. The latter estimates that, 
in a gapless disk, damping by co-orbital Lindblad resonances is about a factor 
of three more rapid than excitation by external Lindblad resonances. In recent 
papers, \cite{WAH98, WAH00} claim that interactions at apsidal resonances, 
which damp eccentricity, are more potent than those at ordinary Lindblad 
resonances. On balance, current opinion favors the notion that planet-disk 
interactions damp eccentricity. We suggest that for massive planets the 
opposite may be true.

The structure of our paper is as follows.  In \S \ref{sec:planet-planet}, we 
argue following \cite{RAF96} and \cite{FHR01} that planet-planet interactions 
cannot account for the prevalence of 
eccentric orbits among isolated planets. \S \ref{sec:planetdiskgen} reviews 
the effects on eccentricity of planet-disk interactions at Lindblad, 
corotation, and apsidal resonances. We demonstrate in \S \ref{sec:apsidal} that 
interactions at apsidal resonances are much less effective at damping 
eccentricity than previous estimates suggest. \S\ref{sec:Lindblad} is devoted 
to an evaluation of the balance between eccentricity damping at Lindblad 
resonances which overlap the planet's orbit and eccentricity excitation at 
those which lie either inside or outside it. A planet which clears a 
sufficiently clean gap tilts the balance in favor of net eccentricity 
excitation by Lindblad resonances. However, linear theory predicts that 
corotation resonances enforce eccentricity damping. This leads to \S 
\ref{sec:saturation} in which we show that the torques at corotation resonances 
are weakened by saturation provided the planet's orbital eccentricity is more 
than a few percent.

In assembling the case that orbital eccentricities of extrasolar planets may 
result from planet-disk interactions, we derive three new technical results.
\begin{itemize}
\item{A correction to the standard Lindblad torque formula for apsidal 
resonances which applies when apsidal waves cannot propagate.}
\item{An account of the balance between eccentricity excitation by Lindblad 
resonances and eccentricity damping by corotation resonances in a gap where the 
pressure gradient reduces the epicyclic frequency of the gas to a value that is 
much smaller than its orbital frequency.} 
\item{A treatment of the saturation of corotation resonances that includes 
the competition between the smoothing of the surface density gradient by the 
mass flux driven by the corotation torque and the action of viscous diffusion 
which acts to reestablish the large scale density gradient.}
\end{itemize}

\section{Planet-Planet Interactions\label{sec:planet-planet}}

Is the observed eccentricity distribution of extrasolar planets the result of
planet-planet interactions? In particular, can they account for the following
properties:
\begin{itemize}
\item {Typical eccentricities are of order a few tenths.} 
\item {Eccentricities smaller than one tenth are rare except for planet's with 
orbital periods less than 20 days where tidal friction is likely to have damped 
eccentricity.}
\item {Most of the planets discovered to date are not currently involved in a 
significant interaction with another planet of comparable or larger mass.}
\end{itemize}

For planet-planet interactions to produce an eccentric orbit for an isolated 
planet requires at least one planet to disappear. The missing planet 
might have collided and merged with the remaining planet or effectively merged 
due to tidal capture. It might also have been 
ejected from the system or fallen into the central star. Numerical integrations 
of systems with two equal mass planets on initially circular orbits by 
\cite{FHR01} produce a much greater fraction of isolated planets with low 
eccentricity orbits than is observed. These are a consequence of mergers which 
\cite{FHR01} assume to occur whenever the separation between the planets drops 
below the sum of their radii. But the case against planet-planet interactions 
is even stronger than the results of \cite{FHR01} indicate because they neglect
tidal captures. Taking the tidal capture cross section for $n=1$ polytropes 
from \cite{KIL99}, and the relative velocity at infinity as numerically 
calculated by \cite{RAF96}, we estimate a critical impact parameter for tidal 
capture between two and three times larger than the two radii assumed for 
merger by \cite{RAF96} and \cite{FHR01}. 

Next we consider some aspects of the merger process for a system whose
initial state consists of two planets, each having mass $M_p$ and radius $R_p$, 
moving on coplaner circular orbits with radii $r_1$ and $r_2$ around a star of 
mass $M_*$ and radius $R_*$. We assume that the final state consists of a 
single or binary planet with mass $2M_p$ moving on an orbit with semimajor axis 
$a$ and eccentricity $e$.\footnote{Orbital elements for a binary refer to its 
center of mass.} Applying conservation of energy and angular momentum, we 
relate the final orbit to the initial ones by 
\begin{equation}
\label{Eb}
E=-{G M_*M_p \over a}=-{G M_* M_p\over 2r_1}-{G M_* M_p\over 2 r_2}-\Delta E ,
\end{equation}
and
\begin{equation}
\label{Hb}
H=2M_p\sqrt{G M_* a(1-e^2)}=M_p\left(\sqrt{G M_* r_1}+\sqrt{G M_* 
r_2}\right)-\Delta H.
\end{equation}
Here $\Delta E$ and $\Delta H$ are the energy and angular momentum stored 
internally in either the merger product or in the relative orbit of the binary. 
Energy dissipated by impact or through tidal dissipation is accounted for by
an increase of the binding energy.  

We estimate $\Delta E$ and $\Delta H$ by noting that when the planets are 
separated by less than the Hill radius, $r_{\rm Hill}\equiv (M_p/3M_*)^{1/3}a$, 
they are effectively a two-body system. Thus\footnote{\cite{FHR01} provide a 
similar derivation to ours. However they incorrectly assume that $\Delta 
E/E\lesssim (M_p/M_*)^{1/2}$ which leads to a higher eccentricity estimate.}
\begin{equation}
{\Delta E\over E} \lesssim {M_p a\over M_* r_{\rm Hill}}\sim \left({M_p\over 
M_*}\right)^{2/3}.
\end{equation}
Moreover, since merging or tidal capture 
requires that the planet-planet periapse distance be no larger than a few times 
$R_p$,
\begin{equation}
\Bigl|{\Delta H\over H}\Bigr|\sim \left({M_pR_p\over M_* 
a}\right)^{1/2}\sim\left({M_p\over 
M_*}\right)^{2/3}\left({R_*\over a}\right)^{1/2}.
\end{equation}
The final expressions for $\Delta E/E$ and $\Delta H/H$ follow from the 
assumption that the planet and star have similar densities. In what follows we 
discard $|\Delta H/ H|\ll \Delta E/E$ since $R_*/a\ll 1$.

Eccentricity is related to orbital energy and angular momentum by 
\begin{equation}
e^2=1+{2H^2E\over (2 M_p)^3  (GM_*)^2}. 
\label{eq:e}
\end{equation}
Substituting for $E$ and $H$ using equations \refnew{Eb} and \refnew{Hb} yields
\begin{equation}
e^2={1 \over 4} \left( 3-{r_1^2 + r_2^2 \over 2r_1r_2}-\sqrt{r_1 \over 
r_2}-\sqrt{r_2 \over r_1} \right)
-  {(r_1+r_2+2\sqrt{r_1r_2}) \Delta E \over 4 GM_*M_p}
\end{equation}
Taking the leading 
contribution in $|r_1-r_2|/(r_1+r_2)$ and writing
\begin{equation}
{\Delta E\over E}=-A\left({M_p\over M_*}\right)^{2/3},
\end{equation}
where $A\lesssim 1$, we deduce that the 
eccentricity is bounded by
\begin{equation}
e^2 = A\left({M_p\over M_*}\right)^{2/3} -{3 \over 4} \left( {r_1-r_2 \over
r_1+r_2}\right)^2.
\end{equation}
Maximal eccentricity is achieved if the initial orbits have the same semimajor 
axis. Mergers or captures are not possible for initial separations
much larger than the Hill radius. 
We conclude that merging and tidal capture cannot produce eccentricities of 
more than a few percents for Jupiter like planets.

Our discussion of planet-planet interactions giving rise to orbital 
eccentricities is far from exhaustive. Here we briefly mention a 
scenario analyzed by \cite{CFT02}. In it two planets undergo differential 
orbital migration which causes the ratio of their mean motions to diverge. 
Passage through mean motion resonances is shown to lead to eccentricity growth.
A shortcoming of this work is that although planet-disk interactions are taken 
to be responsible for orbital migration, their direct effects on eccentricity 
evolution are not accounted for. 

\section{Planet-Disk Interaction - General}
\label{sec:planetdiskgen}

The gravitational potential of a planet can be expanded in a Fourier
series in azimuthal angle $\theta$ and time $t$. Each term of this
series is proportional to $\cos[m(\theta-\Omega_{l,m}t)]$ and has a
radius dependent amplitude $\phi_{l,m}(r)$. The mean motion $\Omega_p$
is the unique pattern speed for a planet with a circular orbit. From
here on, subscripts $p$ and $d$ will denote planet and disk,
respectively. Pattern speeds for a planet which moves on an eccentric
orbit may contain harmonics of the epicyclic frequency $\kappa_p$ and
are denoted by $\Omega_{l,m}=\Omega_p + (l-m)\kappa_p/m$. To first
order in eccentricity $e_p$, each value of $m$ contributes three
components, a principal one with pattern speed $\Omega_{m,m}=\Omega_p$
whose amplitude $\phi_{m,m}$ is independent of $e_p$, and two first
order components with pattern speeds\footnote{The first order $m=0$
terms are an exception. They should be combined into a single
component proportional to $\cos[\kappa_p t]$, formally
implying an infinite pattern speed. More on this is in the discussion section.}
$\Omega_{m\pm 1,m}=\Omega_p \pm \kappa_p/m$ whose amplitudes
$\phi_{m\pm 1,m}$ are proportional to $e_p$.\footnote{In this paper we
consider potential components only up to first order in $e_p$.}

Two kinds of resonance are associated with each potential component.
Corotation resonances occur where the pattern speed matches the angular
velocity of the disk material, $\Omega_{l,m}=\Omega_d$.  A disk particle
located at a corotation resonance experiences a constant torque which causes
the radius of its orbit to change but does not excite its epicyclic motion.
Lindblad resonances occur where the disk's angular velocity differs from the
pattern speed such that $m(\Omega_d-\Omega_{l,m})=\pm\kappa_d$.  The two
Lindblad resonances associated with each potential component are distinguished
by the adjectives inner and outer and are often denoted as $ILR$ and $OLR$.  A
disk particle located at a Lindblad resonance is subject to radial and
azimuthal perturbation forces which vary at its epicyclic frequency.  These
excite its epicyclic motion and also change its semimajor axis.

Each $m$ has nine resonances associated with it: three potential components 
$\phi_{m,m}$ and $\phi_{m\pm 1,m}$, and three resonances for each potential 
component. Table \ref{t:afterglow} describes some properties of these 
resonances.

\begin{center}
\begin{table*}[ht!]
\begin{center}
\begin{tabular}{|c|c|l|c|c|c|c|}
\hline
potential                                 &   pattern speed & 
\multicolumn{2}{c|}{Torque -- $T_d$}  & Keplerian position &  
\multicolumn{2}{c|}{effects} \\ 
                                          & $\Omega_{l,m}$  &     name        & 
   sgn     &                    &      $a_p$      &      $e$      \\  
\hline\hline
                                          &                 & $T^{OLR}_{m,m}$ & 
   $+$     &  $a_p(1+2/3m)$     & \bld\Downarrow  & $\uparrow$  \\      
\raisebox{2.0ex}[0pt]{$\phi_{m,m}$ }      &   $\Omega_p$    & $T^{CR}_{m,m}$  & 
    ?      &  $a_p$             &        ?        &      ?        \\ 
\raisebox{1.5ex}[0pt] {Principal}    &                 & $T^{ILR}_{m,m}$ & 
   $-$     &  $a_p(1-2/3m)$     & \bld\Uparrow    & $\downarrow$    \\ \hline
                                          &                 & 
$T^{OLR}_{m-1,m}$&    $+$    &  $a_p(1+4/3m)$     & $\downarrow$    & 
\bld\Uparrow    \\      
\raisebox{2.0ex}[0pt]{$\phi_{m-1,m}$}     &$\Omega_p-\kappa_p/m$& 
$T^{CR}_{m-1,m}$ & $(-)$ &  $a_p(1+2/3m)$     & $(\uparrow)$    & 
(\bld\Downarrow)\\ 
\raisebox{1.5ex}[0pt] {First Order} &                 & 
$T^{ILR}_{m-1,m}$&    $-$    &  $a_p$             & $\uparrow$      & 
$\downarrow$  \\ \hline
                                          &                 & 
$T^{OLR}_{m+1,m}$&    $+$    &  $a_p$             & $\downarrow$    & 
$\downarrow$  \\      
\raisebox{2.0ex}[0pt]{$\phi_{m+1,m}$}     &$\Omega_p+\kappa_p/m$& 
$T^{CR}_{m+1,m}$ &  $(+)$  &  $a_p(1-2/3m)$     & $(\downarrow)$  & 
(\bld\Downarrow)\\ 
\raisebox{1.5ex}[0pt] {First Order} &                 & $T^{ILR}_{m+1,m}$  
&    $-$    &  $a_p(1-4/3m)$     & $\uparrow$      & \bld\Uparrow    \\ \hline
\end{tabular}
\end{center}
\par
\label{t:afterglow}
\caption{ The nine resonances for a given value of $m$. The sign of
the disk torque, $T_d$, is listed. Up and down arrows denote whether $T_d$ 
increases or decreases $a_p$ and $e_p$. Leading terms, assuming a clean gap in 
a Keplerian disk, are distinguished by 
double arrows. Parentheses denote a dependence on $d(\Sigma/B)/dr$ with the 
sign appropriate to $\Sigma/B$ decreasing towards the planet.}
\end{table*}
\end{center}

\subsection{Understanding the sign of the torque} 
\label{subsec:torquesign}

Each potential component is constant in a frame rotating with its pattern speed
$\Omega_{l,m}$, so its perturbations of the disk's angular momentum $H_d$ 
and energy $E_d$ must preserve the Jacobi constant  
$J_{l,m}=E_d-\Omega_{l,m}H_d$. Thus
\begin{equation}
{dE_d\over dt}=\Omega_{l,m}{dH_d\over dt}=\Omega_{l,m}T_d, 
\label{Jacobi}
\end{equation}
where $T_d$ is the torque exerted on the disk. Provided the unperturbed disk is 
circular, $e_d=0$, equation \refnew{eq:e} applied to the disk material on which 
the torque $T_d$ is applied yields
\begin{equation}
\left(\Omega_{l,m}-\Omega_d\right)T_d={\Omega_d H_d\over 2}{de_d^2\over dt}\ge 
0.
\label{eq:e2Td}
\end{equation}
Equality obtains if the epicyclic motion of the disk material is not 
excited. Thus a potential component may be viewed as transporting angular 
momentum from higher to lower angular velocity \citep{LYK72}. Consequently, 
$T_d>0$ at an OLR and $T_d<0$ at an ILR. 

Because $\Omega_{l,m}-\Omega_d=0$ at a corotation resonance, the sign of the
corotation torque depends upon whether the interaction is dominated by material
inside or outside corotation.  A simple understanding may be obtained by
considering the behavior of collisionless particles.  Disk particles close to
the resonance experience slowly varying torques proportional to
$\sin[m(\theta-\Omega_{l,m}t)]$ which change their orbital radii.  Particles
moving towards corotation drift at a decreasing rate with respect to the phase
of the potential.  The opposite pertains to particles moving away from
corotation.  As a result, there is a flux of particles towards corotation on
both sides of the resonance.  Material outside corotation loses angular
momentum and that inside gains it.  The corotation torque depends upon the
difference and is proportional to $-d(\Sigma/B)/dr$.  Oort's constant
$B=(2r)^{-1}d(r^2\Omega)/dr$, which is equal to $\Omega/4$ for Keplerian
rotation, appears because the radial drift rate is equal to the torque divided
by $2B$.  Pressure effects, neglected here but accounted for later, spread the
region of influence over a radial width which for a Keplerian disk is 
comparable to the disk's vertical scale height.  However, they do not affect 
the net corotation torque.

We note that the corotation torque is negative if material outside the 
resonance dominates the interaction. This might appear surprising in light of 
our finding, based on the Jacobi constant, that torques at Lindblad resonances 
transfer angular momentum from higher to lower angular velocity. Wouldn't a 
similar conclusion apply to corotation resonances?  The answer is no, but for  
a subtle reason. In evaluating the Jacobi constant near a Lindblad resonance 
we neglect the planet's perturbation potential in the energy budget.
That is fine there, but it is not appropriate near a corotation resonance.  

\subsection{Understanding the effects of resonances on the
eccentricities.} 
\label{subsec:edot}

Angular momentum and energy transferred to the disk at a resonance is removed 
from the planet's orbit. Consequently, 
\begin{equation}
{dE_p\over dt}=\Omega_{l,m}{dH_p\over dt}=-\Omega_{l,m}T_d.
\label{Jacobip}
\end{equation}
An alternate derivation proceeds from the fact that the linear density 
perturbation at a resonance rotates with the pattern speed. Thus the perturbed 
disk's gravitational backreaction on the planet preserves the planet's Jacobi 
constant. Combining equations \refnew{eq:e} and \refnew{Jacobip}, we find
\begin{equation}
\label{ededtg}
e_p{de_p \over dt}= 
\left( \left(1-e_p^2\right)^{-1/2}\Omega_p - \Omega_{l,m} \right) 
{H_p^2 T_d\over 
(GM_*)^2M_p^3}.
\end{equation}
At first order resonances $\Omega_{l,m} \ne \Omega_p$, so to lowest order in 
$e_p$,
\begin{equation}
\label{ededte}
e_p{de_p \over dt}= 
\left(\Omega_p -\Omega_{l,m} \right) {H_p^2 T_d\over (GM_*)^2M_p^3}.
\end{equation}
For first order Lindblad resonances, we make use of equation 
\refnew{eq:e2Td} to prove that  
\begin{equation}
{\rm sgn}\left( {de_p \over dt}  \right) = 
{\rm sgn}\left[ (\Omega_p-\Omega_{l,m})(\Omega_{l,m}-\Omega_d) \right].
\label{eq:signedot}
\end{equation}
Thus $e_p$ decreases if the resonance resides on the same side of corotation as 
the planet, and increases if it resides on the opposite side. Henceforth we 
adopt the terminology co-orbital\footnote{In a Keplerian disk these resonances
overlap the planet's semimajor axis.} and external to distinguish these two 
classes of first order Lindblad resonances. 

For first order corotation resonances the sign of the torque is opposite to the 
sign of $d(\Sigma/B)/dr$. Therefore,
\begin{equation}
{\rm sgn}\left({de_p \over dt} \right) = 
{\rm sgn} \left[(\Omega_{l,m}-\Omega_p){d\over dr}\left({\Sigma\over B}\right) 
\right]
\end{equation}
If the planet has cleared a gap, the density increases away from it so both 
first order corotation resonances cause $e_p$ to decay.

Next we examine how the principal Lindblad resonances affect the 
planet's eccentricity. Since $\Omega_{m,m}=\Omega_p$, we retain the lowest 
order dependence on $e_p$ in equation \refnew{ededtg} which reduces to
\begin{equation}
\label{ededtne}
{1\over e_p}{de_p \over dt}= {\Omega_p H_p^2 T_d\over 2(GM_*)^2 M_p^3}.
\end{equation}
\cite{GOT80} compare the rates at which the principal and first order 
Lindblad resonances of the same $m$ change the planet's eccentricity. They find 
that the former is smaller than the latter by a factor $m$, although both have 
the same dependence on $e_p$. We offer some additional comments. Because $T_d$ 
is negative at an ILR and positive at an OLR, the effect of the principal 
Lindblad resonances on $de_p/dt$ suffers from the same cancellation as their 
effect on $da_p/dt$. Rates of change of eccentricity and semimajor axis 
due to principal Lindblad resonances satisfy
\begin{equation}
\label{dedtdadt}
{4\over e_p}{de_p \over dt}= -{1\over a_p}{da_p \over dt}.
\end{equation}
A planet which clears a gap migrates inward as the disk accretes, so its 
eccentricity 
increases. But since the timescale for eccentricity change is four times that 
for radial migration, principal Lindblad resonances are of negligible 
importance in eccentricity evolution.

Now we compare the timescale for eccentricity change, $t_e$, with that for 
semimajor axis migration, $t_{vis}$. The semimajor axis of a planet that opens 
a gap evolves with the disk on the viscous timescale
\begin{equation}
t_{vis}^{-1}\sim {\nu\over r^2}\sim \alpha\Omega\left({h\over r}\right)^2.
\label{eq:tvis}
\end{equation}
Here $\alpha$ is the standard viscosity parameter for accretion disks, and
$h\approx c_s/\Omega$ is the vertical thickness of the disk with $c_s$ the 
sound speed. The planet's orbital eccentricity changes on the timescale
\begin{equation}
t_e^{-1}\equiv {1\over e_p}\Bigl|{de_p\over dt}\Bigr|\approx 
\left({r\over \w}\right)^4\left({M_p\Sigma 
r^2\over M_*^2}\right)\Omega,
\label{eq:te}
\end{equation}
where $\w$ is the gap's width. 
Equation \refnew{eq:te}, adapted from \cite{GOT80}, is obtained by summing 
contributions to 
$de_p/dt$ from first order resonances within a narrow ring separated from the 
planet's position by an empty gap. Assuming Keplerian rotation, they find that 
$e_p$ decays. However damping by corotation resonances exceeds driving by 
Lindblad resonances by only a small, $4.6\%$, margin. The timescale quoted 
above 
for eccentricity change is that due to either corotation or Lindblad resonances 
acting separately. Conditions under which this, rather than the net 
contribution from both types of resonance, is the appropriate $t_e$ to compare 
with $t_{vis}$ are described in \S\ref{sec:saturation}. To elucidate the 
comparison between 
$t_e$ and $t_{vis}$, we relate $\w$ to $\alpha$ and $M_p/M_*$ by balancing
the viscous torque\footnote{As written below, $T_{vis}$ is appropriate to a
Keplerian disk. More generally, the factor of $3$ should be replaced by
$(2B)^{-1}|rd\Omega/dr|$.} 
\begin{equation}
T_{vis}=3\pi\nu\Sigma\Omega r^2=3\pi\alpha\Sigma r^2 (\Omega h)^2, 
\label{eq:Tvis}
\end{equation}
with the torque from the principal Lindblad resonances, 
\begin{equation}
T_L\approx \left({r\over \w}\right)^3 \Sigma r^2(r\Omega)^2\left({M_p\over 
M_*}\right)^2,
\label{eq:TL}
\end{equation}
to obtain
\begin{equation}
{\w\over r}\approx (3\pi\alpha)^{-1/3}\left({r\over h}{M_p\over 
M_*}\right)^{2/3}.
\label{eq:mgap}
\end{equation}
Our typical parameters, $\alpha=10^{-3}$, $h/r=0.04$, $M_p/M_*=10^{-3}$, give 
$\w\approx 0.4 r$.
Substituting this expression for $\w$ into equation \refnew{eq:te} yields
\begin{equation}
t_e^{-1}\approx  {r\over \w}{\Sigma 
r^2\over M_p}t_{vis}^{-1}.
\label{eq:tep}
\end{equation}
For $\w/r\approx 1$, eccentricity evolves faster than semimajor axis provided 
the disk is more massive than the planet.

Co-orbital Lindblad resonances can be ignored in investigations of
the eccentricities of the orbits of shepherd satellites and narrow
planetary rings because the surface density at the orbit of the
shepherd satellite is negligible. But because the ratio of vertical thickness 
to orbital radius is so much larger in protoplanetary disks, $h/r\sim 0.04$, 
than in planetary rings, $h/r\lesssim 10^{-6}$, even a massive planet may not 
clear a very clean gap so interactions at co-orbital Lindblad resonances could 
be significant. As shown by equation \refnew{eq:signedot}, these
interactions damp the planet's orbital eccentricity.  Attempts
to estimate the rate at which first order co-orbital Lindblad
resonances damp eccentricity are described by \cite{WAR88} and
\cite{ART93b}. The latter concludes that, in a disk of uniform surface
density, they damp eccentricity about three times faster than external
first order Lindblad resonances excite it.

An apsidal resonance is a first order inner Lindblad resonance with
$l=0$ and $m=1$. Its pattern speed is equal to the planet's apsidal
precession rate; $\Omega_{0,1}=\Omega_p -\kappa_p =\dot \pomega_p$. As
a co-orbital first order Lindblad resonance, it
contributes to eccentricity damping. Application of the standard
Lindblad resonance torque formula to the apsidal resonance led
\cite{WAH98,WAH00} to conclude that an apsidal resonance damps eccentricity 
much faster than external first order Lindblad resonances excite it.

We have described three types of resonant planet-disk interactions that damp
eccentricity and one that excites it.  In what follows we argue that
eccentricity can grow provided its initial value exceeds some minimal threshold
and the planet clears a sufficiently clean gap.  A minimal initial eccentricity
is necessary for the nonlinear saturation of corotation torques.  This involves
a reduction of the surface density gradient at the resonance position.  A clean
gap renders ineffective damping by first order co-orbital Lindblad resonances.
Although damping due to the apsidal resonance might appear to be the most
potent of all, it is likely to be unimportant.  Apsidal waves have such long
wavelengths in protoplanetary disks that they do not propagate. Thus torques at
apsidal resonances are much smaller than predicted by the standard
Lindblad resonance torque formula.

\section{Apsidal Resonances \label{sec:apsidal}}

Apsidal resonances are special because the doppler shifted forcing frequency 
detunes more gradually from the epicyclic frequency with distance away from an 
apsidal resonance than it does at other Lindblad resonances. This is why the 
standard Lindblad resonance torque formula predicts an apsidal torque  
\begin{equation}
T^{ILR}_{0,1}\sim - e_p^2\left( {M_p \over M_*} \right)^2 {\Sigma 
r^8\Omega^2\over (\Delta r)^4}{\Omega \over r|d\dot\pomega_d/dr|},  
\label{eq:standaps}
\end{equation}
which is larger by $\Omega/{\dot \pomega_d} \sim 10^3$ than that at an ordinary 
first order Lindblad resonance at a distance $\Delta r\lesssim r$ from the 
planet's orbit \citep{GOT78,WAH98,WAH00}. For the same reason, apsidal waves 
have longer wavelengths than those at ordinary Lindblad resonances.  

The long wavelengths of apsidal waves call into question the applicability 
of the standard torque formula. Of particular concern is the assumption that 
density waves propagate away from the resonance and ultimately dissipate 
without reflection. To estimate the first wavelength, $\lambda_1$, of a 
density wave at a Lindblad resonance, we start from the WKBJ dispersion 
relation, $k^2c_s^2=m^2(\Omega_d-\Omega_{l,m})^2-\kappa_d^2\equiv -D$, which 
applies 
in a disk with negligible self-gravity.\footnote{Because of the long 
wavelengths of apsidal waves in protoplanetary disks, a more complete treatment 
should include the disk's self-gravity.} 
We define $\lambda_1$ by
\begin{equation}
\int_{r_*}^{r_*+\lambda_1} k(r)dr=2\pi,
\end{equation} 
where $r_*$ is the resonance radius. Expanding $D\approx {\cal D}(r-r_*)/r_*$ 
around the 
resonance position, we obtain
\begin{equation}
\lambda_1=(3\pi)^{2/3}(c_s^2/|{\cal D}|r_*^2)^{1/3}.
\end{equation}

For apsidal waves $\Omega_{0,1}=\dot\pomega_p$, which implies $D\approx 
2\Omega_d(\dot\pomega_p-\dot\pomega_d)$. Crudely approximating 
$d\dot\pomega_d/dr 
\approx -\dot\pomega_d/r$, we obtain ${\cal D} \approx 
2\Omega_d{\dot\pomega_d}$. 
Hence 
\begin{equation}
\label{lambda1or}
\lambda_1/r \approx  (9\pi^2/2)^{1/3}  (\Omega_d/\dot\pomega_d)^{1/3} 
(h/r)^{2/3}.
\end{equation}
Substituting reasonable parameters for a protoplanetary disk,
$\Omega_d/\dot\pomega_d \sim 10^3$ and $h/r \sim 0.04$, we estimate
that $\lambda_1 \approx 4r$, somewhat longer than the disk radius. 
How does the wavelength far away from the resonance
compare to the local radius?  In the absence of self gravity apsidal
waves propagate toward smaller $r$. We expect that $\dot\pomega_d$
increases inward. Applying the dispersion relation to estimate
$\lambda=2\pi/k$ where $\dot\pomega_d\gg \dot\pomega_p$ yields
\begin{equation} 
\label{lambdaor}
\lambda/r=\sqrt{2}\pi (\Omega_d/\dot\pomega_d)^{1/2} (h/r) \approx 5,
\end{equation}
provided we adopt the same values for $\Omega_d/\dot\pomega_d$ and
$h/r$ used in estimating $\lambda_1$. Our estimates for $\lambda/r$
imply that there are not many wavelengths, probably not even one, in an
apsidal wave train. 

Apsidal waves are not traveling waves subject to the nonlinear steepening and
shock dissipation that is the leading form of damping for density waves
generated at ordinary Lindblad resonances \citep{GOR01}.  Instead, they are
standing waves trapped in a cavity bounded by the resonance radius and the
inner edge of the disk, with viscous dissipation as their principal source of
damping.  Consequently, the standard Lindblad resonance torque formula cannot
be applied to determine the torque at an apsidal resonance in a protoplanetary
disk.  Fortunately, simple physical arguments enable us to estimate the factor
by which the apsidal torque is reduced below the estimate provided by the
standard formula.  Suppose that the apsidal cavity contains $N$ wavelength. 
Comparing the viscous dissipation timescale, $(\nu k^2)^{-1}$, with that for 
propagation across a wavelength, $2\pi/(kv_g)=2\pi\Omega_d/(kc_s)^2$, we deduce 
that the apsidal torque is a factor $2\pi N\alpha$ smaller than the standard 
torque formula implies. Thus for plausible parameters, $N\lesssim 1$
and $\alpha\sim 10^{-3}$, it is even smaller than the torque at other first
order Lindblad resonances.

We have deliberately glossed over several items that merit attention. These are 
briefly covered below. 
\begin{itemize}
\item{The location of the apsidal resonance is uncertain. Factors influencing 
the apsidal precession of the disk include its gravity, the planet's 
gravity, and gas pressure. Only the disk's gravity affects the precession 
of the planet's orbit. Thus we might expect that $\Delta r\sim r$.}
\item{Equation \refnew{eq:standaps} implicitly assumes that 
$\lambda_1\lesssim r$. For $\Delta r<\lambda_1<r$, the prediction of the 
standard torque formula should be multiplied by the factor $F\sim (\Delta 
r/\lambda_1)^2$. Appropriate correction factors if either $\Delta r>r$ or 
$\lambda_1>r$, or both apply, remain to be worked out. }
\item{Then there is the issue of the cavity size measured in wavelengths. Our 
estimate of torque reduction assumes an amplitude for the standing cavity wave 
similar to that which a traveling wave would have in the absence of a 
reflecting boundary. This is an adequate approximation provided $N\lesssim 1$, 
but it would be an underestimate for a cavity tuned near resonance.}
\item{As mentioned previously, a more careful treatment would include the 
disk's self-gravity. However, we can safely state that if self-gravity 
dominates gas pressure in the determination of $\lambda_1$, then $\lambda_1$ 
would be larger than calculated by neglecting self-gravity which would further 
reduce the torque.}

\end{itemize}

\section{Lindblad and Corotation resonances\label{sec:Lindblad}}

First order Lindblad resonances are of two kinds, external ones which excite 
the planet's eccentricity, and co-orbital ones which damp it. In a disk of 
uniform surface density, damping by co-orbital resonances is about a factor of 
three faster than excitation by the external ones \citep{ART93b}. Obviously, 
damping due to co-orbital resonances is weaker if the planet orbits within a 
gap. However, surface density gradients associated with a gap activate 
eccentricity damping by corotation resonances. Can a gap be clean enough to 
prevent eccentricity damping by co-orbital Lindblad resonances yet have 
sufficiently shallow density gradients so as to avoid eccentricity damping by 
corotation  resonances? Although we have not been able to answer this question 
in general, the two examples described below suggest that this is not 
possible.\footnote{In this section we neglect possible saturation of corotation 
resonances.} 

\begin{figure}[t]
\begin{center}
\epsscale{1.1}
\plottwo{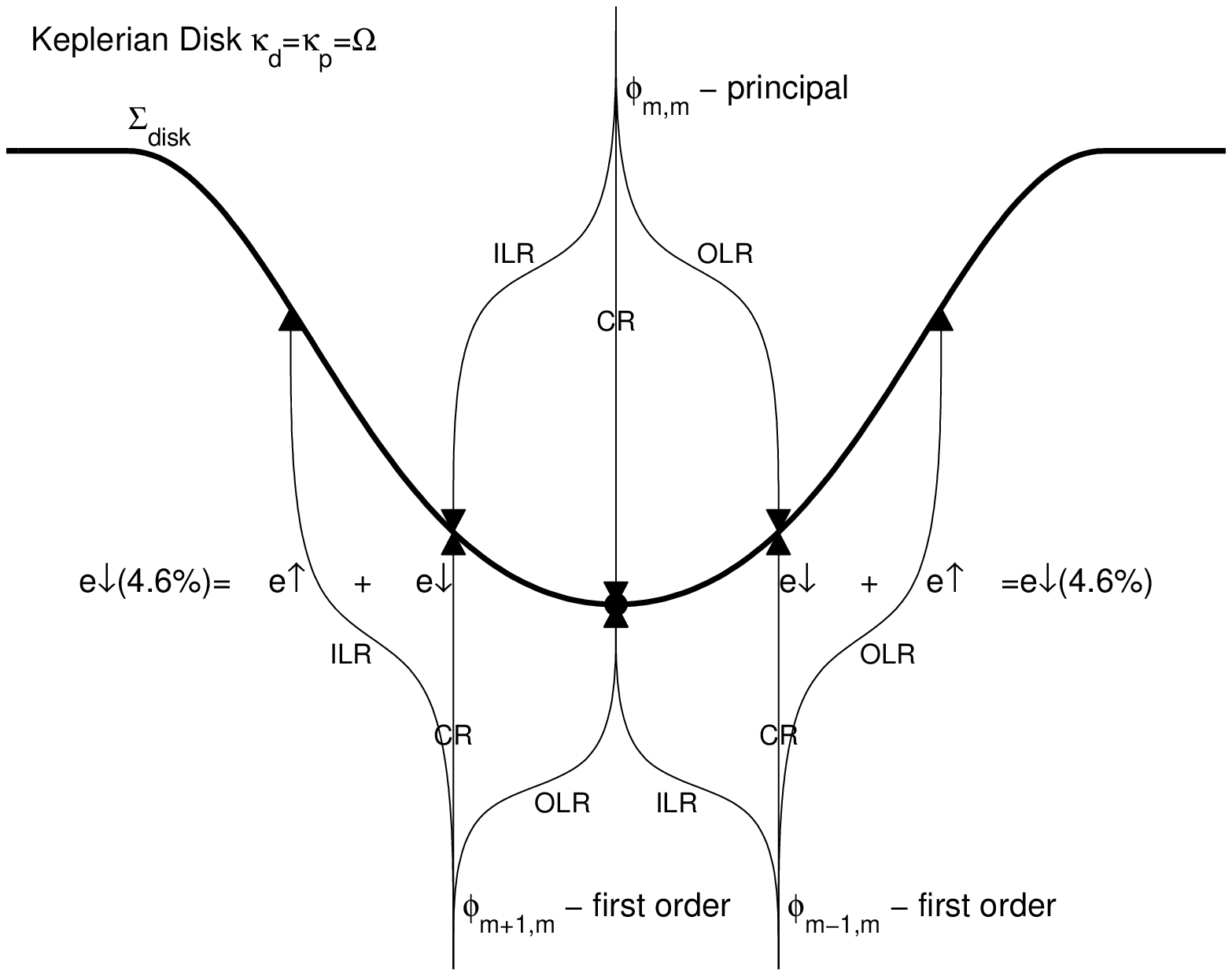}{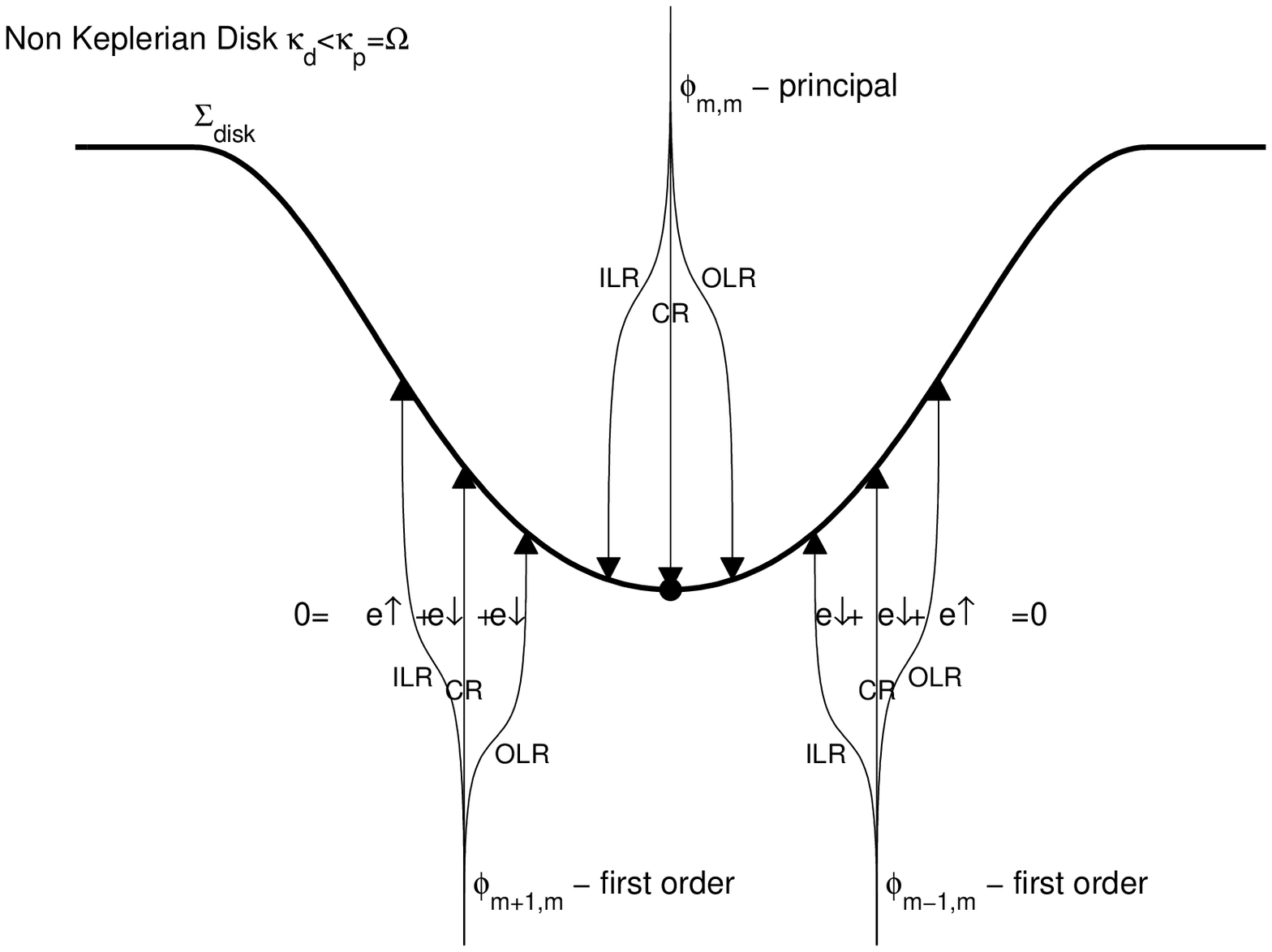}
\end{center}
\caption{ Location of resonances in a Keplerian disk (left), and
in a low epicyclic frequency disk, $\kappa_d/\Omega_d \ll 1$ (right). The 
planet's motion is assumed to be Keplerian, $\kappa_p=\Omega_p$. The smaller 
$\kappa_d/\Omega_d$ is, the closer the principal Lindblad resonances are to the 
gap's center. Competition between eccentricity excitation and damping is tight 
in each case. For a Keplerian disk with a clean gap, the competition is between 
the external Lindblad resonances which excite eccentricity and corotation
resonances which damp it. With decreasing epicyclic frequency, the co-orbital 
and external Lindblad resonances approach each other. Their effects nearly 
cancel leaving a small residual eccentricity growth rate which is more 
completely canceled by the corotation resonances.}
\label{fig1}
\end{figure}

Here we compare the rate of eccentricity damping by co-orbital
Lindblad resonances plus corotation resonances with the rate
of eccentricity excitation by external Lindblad resonances. Our analysis
takes account of a gap in the surface density around the planet's
orbit but assumes that this does not significantly perturb the
Keplerian rotation rate.  For simplicity, we concentrate on the
material external to the planet's orbit.  In this region the external
and co-orbital Lindblad resonances are outer and inner ones
respectively. We employ three constants, ${\cal C}_{eLR}$, ${\cal C}_{cLR}$, 
and ${\cal C}_{CR}$, to describe the relative strengths of
the external Lindblad resonances, the co-orbital Lindblad resonances,
and the corotation resonances. As described in \S
\ref{sec:planetdiskgen}, ${\cal C}_{cLR}/{\cal C}_{eLR}\approx 3$ and
${\cal C}_{CR}/{\cal C}_{eLR}= 1.046$. We take the torque cutoff
to occur at $r_1-a_p\sim h$ for each type of resonance \citep{GOT80, WAR86, 
ART93a}.

Eccentricity excitation by first order external Lindblad resonances is 
evaluated from
\begin{equation}
{1 \over e_p} {de_p \over dt}\Bigr|_{eLR}=
\int_{r_1}^{\infty} {{\cal C}_{eLR} \over (r-a_p)^5} \left( {\Sigma(r) \over 
B_d} \right) dr.
\label{eq:dedtOLR}
\end{equation}
A similar expression describes eccentricity damping by first order 
co-orbital Lindblad resonances, expect that $\Sigma$ is evaluated at the
planet's semimajor axis,
\begin{equation}
{1 \over e_p} {de_p \over dt}\Bigr|_{cLR}
= -\int_{r_1}^{\infty} {{\cal C}_{cLR} \over (r-a_p)^5} \left( {\Sigma(a_p) 
\over B_d} \right) dr
=-{{\cal C}_{cLR} \over 4(r_1-a_p)^4} {\Sigma(a_p) \over B_d}.
\label{eq:dedtILR}
\end{equation}

The effect of corotation resonances on eccentricity evolution is given by 
\begin{equation}
{1 \over e_p} {de_p \over dt}\Bigr|_{CR}=
-\int_{r_1}^{\infty} 
{{\cal C}_{CR}\over 4(r-a_p)^4} {d \over dr} \left( {\Sigma(r) \over B_d} 
\right)  dr.
\end{equation}
Integrating by parts, we arrive at 
\begin{equation}
{1 \over e_p} {de_p \over dt}\Bigr|_{CR}= -\int_{r_1}^{\infty} {{\cal 
C}_{CR}\over (r-a_p)^5} \left( {\Sigma(r) \over B_d} \right) dr + {{\cal 
C}_{CR} 
\over 4(r_1-a)^4} {\Sigma(r_1) \over B_d}. 
\label{eq:dedtCR}
\end{equation}
Except for the negative sign and the substitution of ${\cal C}_{CR}$ for
${\cal C}_{eLR}$, the first term is identical to that for external Lindblad 
resonances. The second term, which is positive, may be viewed as a correction
due to disk material close to the planet's orbit. 

Equations \refnew{eq:dedtOLR}, \refnew{eq:dedtILR}, and \refnew{eq:dedtCR} 
combine to yield
\begin{equation}
{1 \over e_p} {de_p \over dt}= 
 -\int_{r_1}^{\infty}{\left({\cal C}_{CR}-{\cal C}_{eLR}\right)\over (r-a_p)^5}  
\left( {\Sigma(r) \over B_d} \right) dr -{\left({\cal C}_{cLR}\Sigma(a_p)-{\cal 
C}_{CR}\Sigma(r_1)\right)\over 
 4(r_1-a_p)^4B_d}.
\label{eq:dedttot}
\end{equation}
If the planet clears a sufficiently clean gap, only the first term
survives, and eccentricity damping by corotation resonances overcomes
eccentricity driving by external Lindblad resonances, but only by a
small margin since ${\cal C}_{CR}/{\cal C}_{eLR}=1.046$.\footnote{This
situation pertains to narrow planetary rings and their shepherd
satellites \citep{GOT80}.} Residual material in the gap leads to
additional eccentricity damping unless $\Sigma(a_p)/\Sigma(r_1)< {\cal
C}_{CR}/{\cal C}_{cLR} \approx 1/3$. However, such a steep density
gradient near the center of the gap is incompatible with our
assumption that the disk maintains Keplerian rotation.

Next we examine how departures from Keplerian rotation might affect
the balance between eccentricity driving and damping. Gas pressure
perturbs the disk's epicyclic frequency. It decreases $\kappa_d$ at the
outskirts of a gap and increases it around the middle.\footnote{For an
isothermal gas the leading correction to $\kappa_d^2$ is proportional to
the second radial derivative of the logarithm of the surface density.}
Provided the gap is sufficiently clean, only resonances in its outer
parts need be considered. Numerical simulations of gap formation by planets 
show that this limit can be achieved, at least in two-dimensional disks. This
requires $1\ll |d^2\ln\Sigma/d\ln r^2|\lesssim (r/h)^2$. The upper limit is
set by the requirements that principal Lindblad resonances are located away 
from the planet's semimajor axis and by the Rayleigh stability criterion. It 
motivates us to 
investigate the eccentricity evolution of a planet that resides in a gap where 
the epicyclic frequency is much smaller than the orbital angular velocity.  

As $\kappa_d/\Omega_d\to 0$, $d\Omega_d/dr\to -2\Omega_d/r$, and the
positions of the inner and outer Lindblad resonances associated with a
given $\phi_{m-1, m}$ potential component approach, from opposite
sides, the location of the corresponding corotation resonance. This
simplifies comparison of the influences of the different resonances on
eccentricity evolution. Inserting these approximations into the
expressions for the torques at first order resonances given by
\cite{GOT79}, we obtain
\begin{equation}
T_{m-1,m}^{CR}\approx - { m\pi^2 \phi^2_{m-1,m} r \over \kappa_d^2} 
{d\Sigma \over dr},
\end{equation}
and 
\begin{equation}
T_{m-1,m}^{OLR,ILR}\approx \pm {m^2\pi^2 \phi^2_{m-1,m}\Sigma \over
\kappa_d^3/\Omega}.
\end{equation}
Note that in the limit $\kappa_d/\Omega_d\to 0$, each Lindblad torque dominates 
the corotation torque. However, the Lindblad torques have opposite signs and
when summed making use of their separations $\mp r\kappa_d/2m\Omega_d$, we 
discover that
\begin{equation}
T_{m-1,m}^{ILR}+T_{m-1,m}^{OLR}={ m\pi^2 \phi^2_{m-1,m} r \over \kappa_d^2} 
{d\Sigma \over dr}.
\end{equation}
So to leading order, the torques at the inner and outer Lindblad resonances 
just
cancel that at the corotation resonance implying $de_p/dt=0$.

To summarize, neither of our examples yields a prediction of
eccentricity growth. For Keplerian rotation eccentricity damping by
corotation resonances is slightly faster than eccentricity excitation
by external Lindblad resonances.  In the extreme limit of vanishing
epicyclic frequency eccentricity damping by corotation and co-orbital
Lindblad resonances just balances eccentricity driving by external
Lindblad resonances.

\section{Saturation of Corotation Resonances}
\label{sec:saturation}

Linear perturbations of the disk are calculated near discrete resonances. 
Density wave fluxes and disk torques obtained from the linear perturbations
are of second order in the planet's potential. By transporting and depositing 
angular momentum, they cause a secular evolution of the disk. As the disk 
evolves, the linear perturbations, the fluxes, and the torques also change. In 
the particular case of a corotation resonance, the disk torque drives a mass 
flux which flattens the gradient of $\Sigma/B$ thereby
shutting down the torque. This is the phenomenon of resonance saturation.

\subsection{collisionless particle disk} 

It is easy to understand the saturation of the corotation torque in a system of
collisionless particles (see discussion in \S \ref{subsec:torquesign}).  Near a
corotation resonance each particle experiences a slowly varying torque which
causes a slow radial drift of its orbit.  There is a net flux of particles
towards corotation on each side of the resonance.  Trapping of particles around
peaks of the potential occurs within a region of width $\Delta r\approx
(\phi/\Omega^2)^{1/2}$ on timescale $\Delta t\approx (r^2/m\phi)^{1/2}$.
Trapped particles circulate around potential peaks.  Their angular momenta vary
periodically but not secularly.  After trapping occurs the corotation torque
vanishes.  A quantitative discussion is given by \cite{GOT81}.

There is an alternate way to view corotation saturation in a
collisionless particle disk that is more in keeping with what happens
in a gas disk. Trapping kills the gradient $d(\Sigma/B)/dr$ within a
radial interval of width $\Delta r$ around the resonance radius. Thus
the action of the potential changes the disk so that the torque shuts
off. 

\subsection{gas disk} 

The following discussion draws on material in \S IV of
\cite{GOT79}. It also presents new results omitting details of their
derivations. These will be the subject of a future paper.

The corotation torque density, that is, the torque per unit radius near 
corotation ($r=r_{CR}$), is given by
\begin{equation}
{\partial T_d\over \partial r}=-2\pi 
r{\overline{\Sigma_1{\partial\phi_{l,m}\over \partial\theta}}}={ 
T_{l,m}^{CR}\over 2 \ell}\exp{(-|r-r_{CR}|/\ell)},
\label{eq:dTdrCR}
\end{equation}
where the overbar denotes an azimuthal average and the total corotation torque
\begin{equation}
T_{l,m}^{CR}={-m\pi^2 \phi_{l,m}^2\over 2|d\Omega/dr|}{\partial\over \partial 
r}\left({\Sigma\over B}\right)\Bigr|_{CR}.
\label{eq:TCR}
\end{equation}
Note that as a result of gas pressure the torque density is spread
over a radial distance $\ell\equiv c_s/\kappa_d$. In a Keplerian disk, 
$\kappa_d=\Omega_d$ so $\ell=h$. A technical 
comment is in order here. Although the torque density is spread over an 
interval of width $\ell$ around corotation, the gradient $d(\Sigma/B)/dr$ which
appears in equation \refnew{eq:TCR} is to be evaluated at the
resonance radius. That is the meaning of the subscript $CR$ which we
attach to the gradient in this section. In a viscous disk the value of
the corotation torque is determined by the average value of this
gradient within a boundary layer of width
\begin{equation}
\delta_\nu\sim \left({\nu\over m|d\Omega/dr|}\right)^{1/3},
\label{eq:wdef}
\end{equation}
where $\nu$ is the effective kinematic viscosity in the disk. Using the 
standard prescription $\nu=\alpha h^2\Omega$,
\begin{equation}
{\delta_\nu \over \ell}\sim \left({\alpha h^2 r\over m \ell^3}\right)^{1/3}.
\label{eq:walpha}
\end{equation}
Thus $\delta_\nu  \ll \ell$ provided $\alpha\ll m \ell^3/(h^2r)$, which we 
assume 
to be true in what follows. 
 
The corotation torque density drives an azimuthally averaged radial
mass flux
\begin{equation}
F_\Sigma \equiv \overline{\Sigma v_r}={1 \over 2Br}\left[ {1\over 2\pi
r}{\partial T_d\over \partial r} + {1\over r}{\partial \over \partial
r}\left(r^3\nu \Sigma {d\Omega\over dr}\right) \right].
\label{eq:FSig}
\end{equation}
Equation \refnew{eq:FSig} has a simple physical interpretation. Each term in 
the square 
brackets is a torque per unit area, the first comes from the planet and the 
second from the viscous stress. Orbits of disk particles drift adiabatically in 
response to absorbing angular momentum at rate inversely proportional to $2Br 
\equiv d(r^2\Omega)/dr$, the radial gradient of specific angular momentum.
However, a rigorous derivation is more complicated since the Reynold's stress 
also transports angular momentum and angular velocity perturbations affect the 
angular momentum density. These two effects cancel each other leaving 
equation \refnew{eq:FSig} intact.

Next we substitute the expression for the mass flux given by equation 
\refnew{eq:FSig} into the equation of mass conservation and use equations 
\refnew{eq:dTdrCR} and \refnew{eq:TCR} to eliminate the corotation torque 
density. Under the assumption that $\Sigma$ varies on a shorter
spatial scale than $\nu$, $\Omega$, $B$, and $d\Omega/dr$, we arrive at
\begin{equation}
{\partial\Sigma\over \partial t}={\nu r|d\Omega/dr|\over 
2B}{\partial^2\Sigma\over \partial r^2} -{\rm sgn}\left(r-r_{CR}\right) {\pi m 
\phi_{l,m}^2\over 
16r^2\ell^2|d\Omega/dr|B^2}{\partial \Sigma\over \partial 
r}\Bigr|_{CR}\exp{(-|r-r_{CR}|/\ell)}.
\label{eq:Sigdot}
\end{equation}
Specializing to near Keplerian rotation, we arrive at
\begin{equation}
{\partial\Sigma\over \partial t}=3\nu{\partial^2\Sigma\over \partial r^2} 
-{\rm sgn}\left(r-r_{CR}\right) 
{2\pi m\phi_{l,m}^2\over 3rh^2\Omega^3}{\partial \Sigma\over \partial 
r}\Bigr|_{CR}\exp{(-|r-r_{CR}|/h)}.
\label{eq:Sigdotkep}
\end{equation}
The large scale density gradient in a gap is maintained by a balance between 
torques from principal Lindblad resonances and the viscous stress. 
Understanding the action of the two terms on the right hand sides of equations 
\refnew{eq:Sigdot} and \refnew{eq:Sigdotkep} is straightforward. The mass flux 
driven by 
the differential corotation torque tends to decrease the density gradient 
within a region of width $\ell$ around corotation. Viscous diffusion acts to 
maintain the density gradient at the value it has on larger scales. For 
parameters of interest to us, the density gradient suffers only a small 
fractional reduction. By integrating the steady state version of equation 
\refnew{eq:Sigdotkep} once with respect to $r$, we find
\begin{equation}
{\partial\Sigma\over \partial r}= 
{\partial\Sigma\over\partial r}\Bigr|_\infty 
-{2\pi m \phi^2_{l,m}\over 9\alpha rh^3\Omega^4}
{\partial\Sigma\over \partial r}\Bigr|_{CR}\exp{(-|r-r_{CR}|/h)}.
\label{eq:DelSig}
\end{equation}
A more revealing form for first order corotation resonances is obtained using 
$\phi_{m\pm 1,m}\sim
me(M_p/M_*)(r\Omega)^2$, $m\approx r/\w$, and substituting equation 
\refnew{eq:mgap}
into equation \refnew{eq:DelSig};
\begin{equation}
{\partial\Sigma\over \partial r}- {\partial\Sigma\over\partial 
r}\Bigr|_\infty \approx -e_p^2{r\over h}{\partial\Sigma\over \partial 
r}\Bigr|_{CR}\exp{(-|r-r_{CR}|/h)},
\label{eq:Sigred}
\end{equation}
which implies
\begin{equation}
{\partial\Sigma\over \partial r}\Bigr|_{CR}\approx 
{1\over 1+e_p^2r/h}{\partial\Sigma\over\partial r}\Bigr|_\infty
\end{equation}
A more careful derivation establishes that equation \refnew{eq:Sigred}
is not missing a significant numerical constant. With $h/r\approx 0.04$,
$e_p\approx 0.2$ is required for complete saturation, but $e_p\approx 0.05$ 
suffices to tip the balance in favor of eccentricity driving by Lindblad 
resonances as opposed to eccentricity damping by corotation resonances.

\begin{figure}[tbp]
\begin{center}
\epsscale{1} \plotone{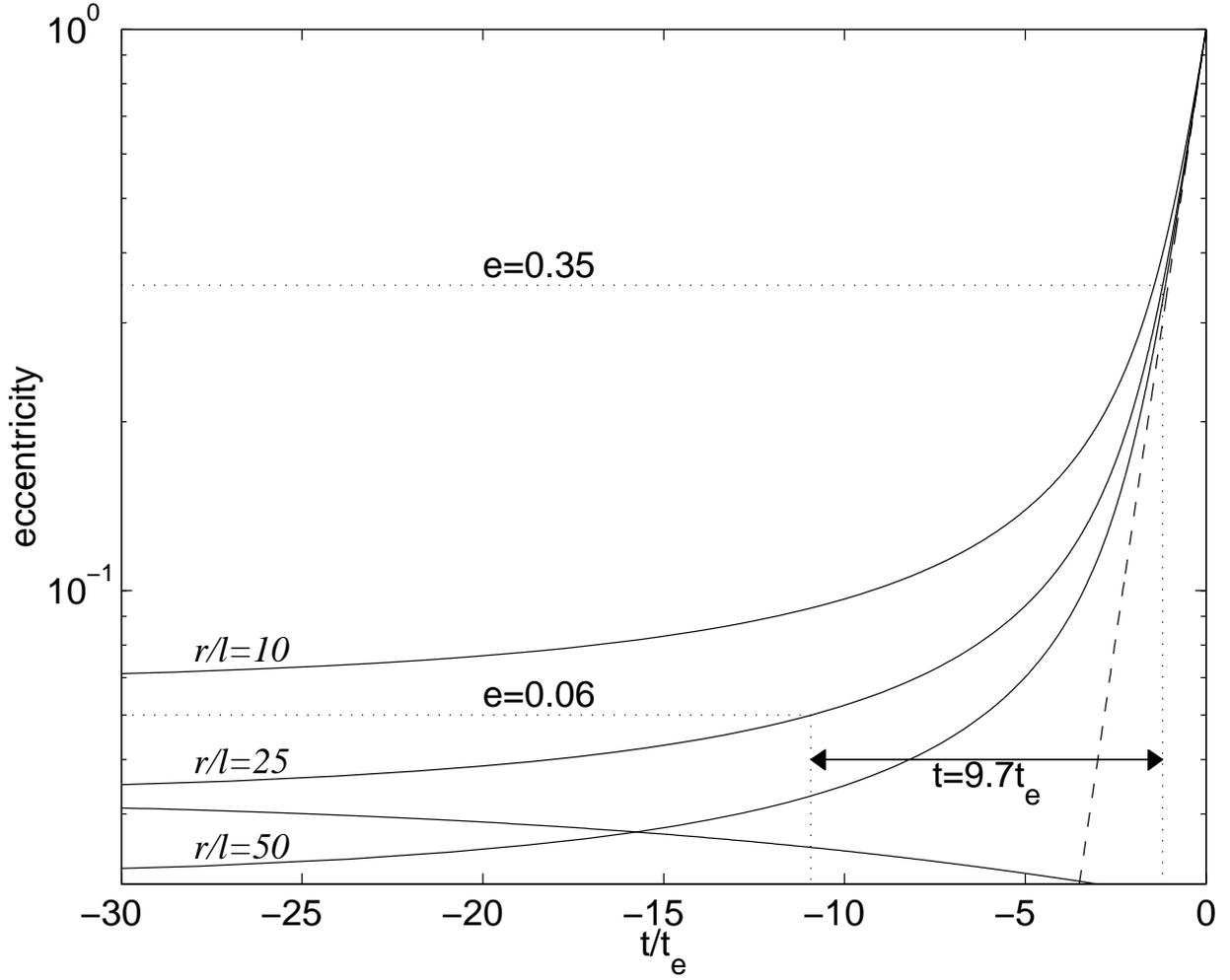}
\end{center}
\caption{Eccentricity evolution with a partially saturated corotation 
resonance. The arbitrary time origin is chosen such that $e=1$ at $t=0$. 
Solid lines show how the eccentricity grows for three values of 
$r/\ell=10,25,50$ 
with ${\cal C}_{CR}/{\cal C}_{eLR}=1.046$. These are to be compared to the 
dashed line which
illustrates eccentricity amplification under complete saturation. For 
$r/\ell=25$ we 
also include a case illustrating the decay of a subcritical 
eccentricity. Dotted lines demonstrate that for $r/\ell=25$ and partial 
saturation 
it takes about ten Lindblad growth timescales for the eccentricity to increase 
from $e=0.06$ to $e=0.35$, whereas for complete saturation it would have taken 
only 1.76.}
\label{fig2}
\end{figure}

Partial saturation of corotation resonances leading to eccentricity growth
further requires that the initial value of $e_p$ be able to survive damping 
until the torque is adequately saturated. To assess the condition under which 
this is satisfied, we compare the timescale for eccentricity damping given by
equation \refnew{eq:te},
\begin{equation}
t_e\sim \left({\w\over r}\right)^4{M_*^2\over M_p\Sigma r^2\Omega},
\end{equation}
with that during which $d\Sigma/dr|_{CR}$ achieves a steady state,
\begin{equation}
t_{sat}\sim {h^2\over \nu }\sim {1\over \alpha \Omega}.
\end{equation}
Applying equation \refnew{eq:mgap} yields
\begin{equation}
{t_{sat}\over t_e}\sim {h^2\over r\w}{\Sigma r^2\over M_p}.
\end{equation}
For plausible parameters, $\w/r\approx 1$ and $h/r\approx 0.04$, this ratio
is less than unity except for very large values of $\Sigma
r^2/M_p$. Under the dual assumptions of a steady state gap and steady state 
saturation, eccentricity growth is governed by
\begin{equation}
{1 \over e_p}{de_p \over dt}={1 \over t_e}\left(1-{{\cal C}_{CR}/{\cal 
C}_{eLR} \over 
1+re_p^2/h}\right)
\end{equation}
Sample solutions of this equation are plotted in figure \ref{fig2}.

\section{Discussion}

Resonant interactions between a planet and a protostellar disk cause the 
planet's orbital eccentricity to evolve on a short timescale. We have 
investigated whether these interactions might be responsible for the eccentric 
orbits that characterize recently discovered extrasolar planets. Our 
preliminary findings are as follows.

Co-orbital Lindblad resonances damp eccentricity.  They dominate the
eccentricity evolution of planets that are too small to clear gaps
\citep{WAR88, ART93b}.  External Linblad resonances excite eccentricity
\citep{GOT80}.  Lindblad resonances, by themselves, lead to net eccentricity
excitation for planets which open clean gaps.  However, gradients of surface 
density are enhanced in the walls of gaps and these activate eccentricity
damping by corotation resonances.  Unsaturated, corotation resonances damp
eccentricity slightly faster than external Lindblad resonances excite it
\citep{GOT80}.  Corotation resonances can saturate because the planet's torque
drives a mass flux which tends to smooth the surface density gradient near
corotation.  However, saturation is only effective if the smoothing occurs on
the timescale of viscous diffusion across a radial scale comparable to the disk
thickness.  This requires the planet to obtain an initial orbital eccentricity
of a few percent.  Thus eccentricity growth due to planet-disk interactions is
a finite amplitude instability.

An apsidal resonance is a special kind of inner Lindblad resonance whose 
pattern speed is equal to the apsidal precession rate of the planet. 
Apsidal waves are longer than density waves excited at ordinary Lindblad 
resonances. The standard Lindblad resonance torque formula predicts 
that the apsidal torque is enhanced relative to that at other Lindblad 
resonances. Formally, this implies that eccentricity damping due to apsidal 
waves is more potent than other forms of eccentricity change due to planet-disk 
interactions. However, because of their long wavelengths, apsidal 
waves are standing rather than traveling disturbances in protoplanetary disks. 
Unlike traveling waves, they do not steepen into shocks and dissipate, but 
instead are weakly damped by viscosity. As a result, the apsidal torque is much 
smaller than the standard torque formula predicts. It is plausible that apsidal 
waves do not play a significant role in the eccentricity evolution of 
extrasolar planets. 

Next we provide concrete example of a system in which eccentricity growth might 
occur. Suppose that a Jupiter mass planet with $M_p/M_* \approx 10^{-3}$, 
initial semimajor axis $a_p\approx\,$AU, and initial eccentricity $e_p \approx 
0.06$, is embedded in a protostellar disk with radius to scale high ratio 
$r/h\approx 25$ and viscosity parameter $\alpha \approx 10^{-3}$. The planet 
would open a wide gap with $\w \approx r/2$. If the gap were as clean as the 
Rayleigh stability condition, $\kappa^2>0$, permits, the residual surface 
density at its center would be so small that eccentricity damping due to 
co-orbital Lindblad resonances would be negligible. Acting separately, either 
eccentricity damping by corotation resonances or eccentricity excitation by 
external Lindblad resonances would proceed on a timescale of order $10^{5}$ 
years. However, if the corotation resonances were unsaturated and they were 
acting together with the Lindblad resonances, eccentricity would damp on a 
significantly longer timescale. Given the initial eccentricity, partial 
saturation of the corotation resonances leading to eccentricity growth would 
occur on a timescale of order $2 \times 10^3$ years, the viscous diffusion 
timescale across a radial interval comparable to the disk's vertical 
scaleheight. As the eccentricity increased, so would the saturation of the 
corotation resonances, and thus the eccentricity growth rate. Figure \ref{fig2} 
shows that the eccentricity could reach the observed mean value of $e=0.35$ 
within about $10^6$ years.

Investigations of eccentricity evolution due to planet-disk interactions are 
plagued by several major uncertainties. 
\begin{itemize}

\item {Angular momentum transport by internal stresses plays a major
role in disk accretion, gap formation, and the saturation of
corotation resonances.  Presumably it arises from an instability, but
we lack a basic understanding of its origin and character. At present,
all we can do is assume that the disk possesses a kinematic $\alpha$
viscosity. Dissipation of disturbances near resonances occurs through
shocks and turbulence. How these contribute to angular momentum
transport is a subject for future investigation.  Analogous phenomena
are involved in maintaining sharp edges in planetary rings
\citep{BGT82a, BGT82b}. }

\item {Clean gaps are essential for the suppression of eccentricity
damping by co-orbital Lindblad resonances. Two dimensional simulations
produce gaps of the requisite cleanliness, but it is not yet known
whether three dimensional simulations will confirm their reality
\citep{BRL00}. The Rayleigh stability criterion, $\kappa^2>0$, limits
the magnitude of density gradients in gap walls. There are also
weaker, nonaxisymmetric instabilities that come into play at positive
$\kappa^2$.  Examining their role in determining gap shape is a task
best suited for a coordinated attack by analytical calculations and
numerical simulations.}

\item {Jovian mass planets should be able to open gaps in protostellar
disks. Those that were not accreted onto and consumed by their central
stars must have been present while the disk dispersed. During the
latter stages of disk dispersal, they would have been more massive
than the disk. It is known from studies of narrow planetary rings that
the growth and damping of a ring's eccentricity is in most respects
identical to that of the satellites which shepherd it
\citep{GOT80}. The ratio of the rates of eccentricity change of ring
and satellite are inversely proportional to the ratio of ring to
satellite mass. To maintain an eccentric shape a narrow ring must
precess rigidly. This requires the wavelength of the apsidal wave with
pattern speed equal to the precessional angular velocity to be longer
than the ring's radial width.  Since apsidal waves in protostellar
disks have wavelengths comparable to the local radius, large regions
of these disks might maintain eccentric shapes.}

\item {Our proposal for eccentricity growth due to planet-disk
interactions involves a finite amplitude instability. However, we do
not have an obvious candidate for giving a planet the requisite
initial eccentricity of a few percent. Gravitational interactions with
surface density perturbations created by instabilities in the walls of
gaps is one possibility. }

\item {Once an adequate core has been assembled, the accretion of gas
to form a Jovian mass planet proceeds rapidly. As its mass increases,
a planet begins to clear a gap which hinders its ability to accrete
additional gas. Eccentricity growth could occur prior to gap attaining
its equilibrium width provided it become sufficiently clean. This
would have two positive effects on the planet's potential for
eccentricity growth. It would lessen the initial eccentricity required
for adequate saturation of corotation resonances since
$e_{crit}\propto \w^{3/2}$ for fixed viscosity, and it would speed up
the rate of eccentricity increase since $de_p/dt\propto \w^{-4}$.}

\item {Saturation of corotation resonances involves the flattening of
$\Sigma/B$ in regions of width $\ell$ around each
resonance. Neighboring resonances are separated by distances of order
$r/m^2$, so these regions overlap for $m>(r/\ell)^{1/2}$. Where
overlap occurs, the interplay between gap maintenance and corotation
saturation will be more complicated than we have described. Presumably
$\Sigma/B$ would flatten in a region narrower than $\ell$ and this
would require a larger eccentricity to overcome viscous diffusion.
Although this is an unresolved difficulty, it is unlikely to be of
great significance since with our standard parameters the most
important resonances lie well outside the region of overlap.}

\item {The sites of first order corotation resonances coincide with
those of principal Lindblad resonances. Linear perturbations
associated with the latter are larger than those of the former by a
factor of order $(me_p)^{-1}$. Whether this affects saturation of
corotation resonances is an open issue.}

\item {We only consider principal and first order resonances in this
paper.  While this is adequate for examining the excitation and
damping of small eccentricities, higher order resonances will have to
be taken into account if we want to examine the growth of eccentricity
to the large values characteristic of extrasolar planets.}

\item {Hill radii of Jupiter mass planets are larger than the vertical
scale heights of protostellar disks. Thus they might at least
temporarily trap gas in quasi two dimensional horseshoe orbits. Then
some of the angular momentum deposited at co-orbital Lindblad
resonances could find its way back into the planet's orbit. This would
reduce the eccentricity damping rate of co-orbital Lindblad
resonances.}

\item {Our analysis is suitable for $m\gg 1$. However, for Jupiter
mass planets, and for $\alpha=10^{-3}$, the size of the gap $\w
\approx r$ suggests $m\approx 1$. Significant corrections may apply in
this limit.}

\item{The first order axisymmetric, $m=0$, resonance deserves special
attention.  It can be thought of as a co-orbital Lindblad resonance, and as the 
arguments of \S\ref{sec:planetdiskgen} demonstrate, it leads to eccentricity 
damping. However, preliminary investigation shows that it is of little 
importance.}

\end{itemize}

Our mechanism of eccentricity growth through saturation of corotation
resonances can ultimately be tested by numerical simulations.
Recently, \cite{PNM01} observed eccentricity growth for planets larger
than about $10-20 M_J$ in an $\alpha=4.5 \times 10^{-3}$ disk. With
these parameters the gap extends past the $2:1$ resonance making all
first order corotation resonances impotent. Similar behaviour was
observed earlier by \cite{ACL91} for binary stars surrounded by a
disk. Both papers appear to support our claim that the apsidal
resonance does not damp eccentricity as fast as external Lindblad
resonaces excite it.  In apparent contradiction with our conclusions,
\cite{PNM01} did not obtain eccentricity growth for Jovian mass
planets. Several explanations come to mind.  Limitations of
our analysis may have led us to predict eccentricity growth where it does
not occur. Alternatively, the simulations may lack sufficient resolution to
capture the partial saturation of corotation resonances. Indeed the grid
spacing of  their Jovian mass simulation was of order $\ell$. Finally,
their initial $e_p$ may be below the limit required for adequate saturation.

\acknowledgements

This research was supported in part by an NSF grant awarded to PG and a Sherman
Fairchild Fellowship held by RS. We thank Geoff Bryden for several illuminating
conversations. 


\end{document}